\def\ra{\rangle}
\def\la{\langle}
\def\be{\begin{equation}}

\def\ee{\end{equation}}
\def\ba{\begin{array}}
\def\ea{\end{array}}

\def\l{\lambda}

\documentclass[aps,pra,showpacs,showkeys,preprint]{revtex4}
\usepackage{amsmath}
\usepackage{amssymb}
\usepackage{graphicx}
\begin{document}

\title{Bell Inequalities Classifying Bi-separable Three-qubit States}

\author{Bao-Zhi Sun$^{a,b}$}
\author{Shao-Ming Fei$^{a,c}$}

\affiliation{$~^{a}$ Department of Mathematics, QuFu Normal University, Shandong 273165\\
$~^{b}$ Department of Mathematics, Capital Normal University, Beijing 100037\\
$~^{c}$ Institut f\"ur Angewandte Mathematik, Universit\"at Bonn, D-53115}

\begin{abstract}
We present a set of Bell inequalities
that gives rise to a finer classification of the entanglement
for tripartite systems. These inequalities distinguish three
possible bi-separable entanglements for three-qubit states.
The three Bell operators we employed
constitute an external sphere of the separable cube.
\end{abstract}

%\date{\today}
\maketitle
\vspace{3ex}

The Bell inequality \cite{Bell64}
provided the first possibility to distinguish experimentally
between quantum-mechanical predictions and
those of local realistic models.
Since Bell's work, there were many important generalizations
\cite{MABK,WW2001,ZB2002,Generalized}.
We refer to \cite{reviews} and references therein as recent nice reviews.

The inequalities in \cite{classification,classificationJ}
can lead to a detailed classification of multipartite
entanglement. By generalizing the method used in \cite{Chen2006}, where
Bell inequalities that violate the generalized GHZ state and
involve only two measurement settings per observer are investigated,
in this letter we present a set of Bell inequalities
that gives rise to a finer classification of the entanglement
for tripartite systems. In classifying bi-separable
three-qubit states, these inequalities determine
further which qubit is separated from the rest two.
Moreover it is shown that the three Bell operators we employed
constitute a sphere. The detailed classifications are depicted
according to parameter regions in the sphere.

Consider $N$ parties and allow each of them to
choose independently between two dichotomic observables $A_{j}$, $B_{j}$
for the $j$-th observer, where $A_j=\vec{a}_j\cdot\vec{\sigma}^j$ and
$B_j=\vec{b}_j\cdot\vec{\sigma}^j$, with $\vec{\sigma}^j=({\sigma_1}^j,{\sigma_2}^j,{\sigma_3}^j)$ the
Pauli matrices on the $j$-th qubit, and $\vec{a}_j=(a^1_j,a^2_j,a^3_j)$,
$\vec{b}_j=(b^1_j,b^2_j,b^3_j)$ the real unit vectors.
We define
\begin{equation}\label{sun}
\mathcal{D}_N^{(i)}=\mathcal{B}_{N-1}^{(i)}\otimes\frac{1}{2}(A_i+B_i)
+\frac{1}{2}(A_i-B_i),\ i=1,\cdots,N,
\end{equation}
where $\mathcal{B}_{N-1}^{(i)}$ is the quantum mechanical Bell operator of
WWZB inequalities \cite{WW2001,ZB2002} on the $N-1$ qubits
except for the $i$-th qubit.

For tripartite case ($N=3$), for example, we have
$\mathcal{B}_2^{(2)}=\frac{1}{2}(A_1B_3+B_1A_3+A_1A_3-B_1B_3)$, and
\begin{eqnarray*}
\mathcal{D}_3^{(2)}&=&\mathcal{B}_2^{(2)}\otimes\frac{1}{2}(A_2+B_2)
+\frac{1}{2}(A_2-B_2)\\
&=&\frac{1}{4}(A_1(A_2+B_2)A_3+A_1(A_2+B_2)B_3+B_1(A_2+B_2)A_3-B_1(A_2+B_2)B_3)\\
&&\ \ +I\otimes\frac{1}{2}(A_2-B_2)\otimes I.
\end{eqnarray*}
It is straightforward to prove that for fully
separable states $\rho$, the average values $\la\mathcal{D}_3^{(i)}\ra_\rho$
of $\mathcal{D}_3^{(i)}$ satisfy
$|\la\mathcal{D}_3^{(i)}\ra_\rho|\leq1$ for $i=1,2,3$.

Let $S_{1-23}$, $S_{2-13}$ and $S_{12-3}$ denote
the bi-separable states of
the form $\rho_1\otimes\rho_{23},\ \rho_2\otimes\rho_{13},$ and
$\rho_{12}\otimes\rho_3$ respectively.  We have

{\sf [Theorem 1]} For states $\rho$ in $S_{1-23}$,
$S_{2-13}$ and $S_{12-3}$, respectively we have:
\begin{equation}\label{sun1}
|\la\mathcal{D}_3^{(1)}\ra_\rho|\leq\sqrt{2},\ \
|\la\mathcal{D}_3^{(2)}\ra_\rho|\leq1,\ \
|\la\mathcal{D}_3^{(3)}\ra_\rho|\leq1,
\end{equation}
\begin{equation}\label{sun2}
|\la\mathcal{D}_3^{(1)}\ra_\rho|\leq1,\ \
|\la\mathcal{D}_3^{(2)}\ra_\rho|\leq\sqrt{2},\ \
|\la\mathcal{D}_3^{(3)}\ra_\rho|\leq1
\end{equation}
and
\begin{equation}\label{sun3}
|\la\mathcal{D}_3^{(1)}\ra_\rho|\leq1,\ \
|\la\mathcal{D}_3^{(2)}\ra_\rho|\leq1,\ \
|\la\mathcal{D}_3^{(3)}\ra_\rho|\leq\sqrt{2}.
\end{equation}

{\sf [Proof]} That $|\la\mathcal{D}_3^{(i)}\ra_\rho|$ has the bound $\sqrt{2}$ for all
3-qubit states can be seen from $\la\mathcal{D}_3^{(1)}\ra^2=\frac{1}{2}(1+\vec{a}_1\cdot\vec{b}_1)
\la\mathcal{B}_3^{(1)}\ra^2+\frac{1}{2}(1-\vec{a}_1\cdot\vec{b}_1)\leq2$, taking into account
the result $\la\mathcal{B}_3^{(1)}\ra^2\leq2$ in \cite{WW2001}.
As an example we consider the states in $S_{12-3}$ and prove the inequalities in (\ref{sun3})
in the following. Due to the linear property of average values, we only need to discuss
pure states. From the Schmidt biorthogonal decomposition theorem
\cite{Nielsen}, every pure state in $S_{12-3}$ can be written as
\begin{equation}\label{psi}
|\psi\ra=(\mbox{cos}\alpha|01\ra-\mbox{sin}\alpha|10\ra)\otimes|0\ra
\equiv|\psi\ra_{12}\otimes|\psi\ra_{3}.
\end{equation}
Therefore
\begin{eqnarray*}
&&|\la\mathcal{D}_3^{(1)}\ra_{|\psi\ra}|\\
&=&|\la\frac{A_1+B_1}{2}\frac{A_2+B_2}{2}\ra_{|\psi\ra_{12}}\la
A_3\ra_{|\psi\ra_3}
+\la\frac{A_1+B_1}{2}\frac{A_2-B_2}{2}\ra_{|\psi\ra_{12}}\la
B_3\ra_{|\psi\ra_3}
+\la\frac{A_1-B_1}{2}\ra_{|\psi\ra}|\\
&\leq&\mbox{sup}|\la\frac{A_1+B_1}{2}\frac{A_2+B_2}{2}\ra_{|\psi\ra_{12}}
+\la\frac{A_1+B_1}{2}\frac{A_2-B_2}{2}\ra_{|\psi\ra_{12}}
+\la\frac{A_1-B_1}{2}\ra_{|\psi\ra}|\\
&=&\mbox{sup}|\la\frac{A_1+B_1}{2}A_2\ra_{|\psi\ra_{12}}
+\la\frac{A_1-B_1}{2}\ra_{|\psi\ra}|\\
&=&\frac{1}{2}\,\,\mbox{sup}|(a_1^3+b_1^3)a_2^3
+\mbox{sin}2\alpha((a_1^1+b_1^1)a_2^1+(a_1^2+b_1^2)a_2^2)+\mbox{cos}2\alpha(a_1^3-b_1^3)|.
\end{eqnarray*}
The maximum is obtained at either
$\alpha=0$ or $\alpha=\frac{\pi}{4}$. For the case $\alpha=0$
the state (\ref{psi}) is
factorizable and the inequality is trivially satisfied.
For the case $\alpha=\frac{\pi}{4}$ we have
$$
|\la\mathcal{D}_3^{(1)}\ra_{|\psi\ra}|\leq
\mbox{sup}\frac{1}{2}|(\vec{a}_1+\vec{b}_1)\cdot\vec{a}_2|=1.
$$
Similarly we can prove $|\la\mathcal{D}_3^{(2)}\ra_{|\psi\ra}|\leq1$.
The inequalities in (\ref{sun1}) and (\ref{sun2}) can be also
proved accordingly. \hfill  $\square$

If we consider $\la\mathcal{D}_3^{(i)}\ra_\rho,\
i=1,2,3$ to be three coordinates, then
all the fully separable states are confined in a cube with size
$2\times 2\times 2$. While from the Theorem 1, the
bi-separable states are in a cuboid with the size either
$2\sqrt{2}\times 2\times 2$ or $2\times 2\sqrt{2}\times 2$
or $2\times 2\times 2\sqrt{2}$, which correspond to
three bi-separable cases: $S_{1-23}$, $S_{2-13}$ and $S_{12-3}$.
The states in other regions are then tripartite entangled.
However there exists a quadratic inequality that strengthens the
range.

{\sf [Theorem 2]} For all 3-qubit states, we have the following
inequality:
\begin{equation}\label{forall}
\la\mathcal{D}_3^{(1)}\ra_\rho^2+\la\mathcal{D}_3^{(2)}\ra_\rho^2
+\la\mathcal{D}_3^{(3)}\ra_\rho^2\leq3.
\end{equation}

{\sf [Proof]} A 3-qubit state $\rho$ can be generally expressed as
\cite{Linden}: \be\label{Pauli}
\rho=\frac{1}{8}(I+\alpha_i\sigma_i^1+\beta_i\sigma_i^2+\gamma_i\sigma_i^3
+R_{ij}\sigma_i^1\sigma_j^2+S_{ij}\sigma_i^1\sigma_j^3+T_{ij}\sigma_i^2\sigma_j^3
+Q_{ijk}\sigma_i^1\sigma_j^2\sigma_k^3), \ee where $\sigma_i^j$
are the $i$-th Pauli matrices on the $j$-th qubit, e.g.
$\sigma_i^1=\sigma_i\otimes I\otimes I$, $\alpha_i$, $\beta_i$,
$\gamma_i$, $R_{ij}$, $S_{ij}$, $T_{ij}$ and $Q_{ijk}$ are some
real coefficients. The repeated indices are assumed to be summed
over from 1 to 3. As the expression
$\omega=\la\mathcal{D}_3^{(1)}\ra_\rho^2
+\la\mathcal{D}_3^{(2)}\ra_\rho^2
+\la\mathcal{D}_3^{(3)}\ra_\rho^2$ is a convex function of $\rho$,
it is sufficient to consider pure states only. Let
$\rho=|\Psi\ra\la\Psi|$. For three qubits states, $|\Psi\ra$ has
the following decomposition \cite{acin}:
\be\label{Psi4}
|\Psi\ra=\l_0|000\ra+\l_1e^{i\phi}|100\ra +\l_2|101\ra+
\l_3|110\ra+\l_4|111\ra,
\ee
with normalization condition
$$
0 \leq \phi \leq \pi,\ \ \lambda_i \geq 0,\ \ \sum_i
\lambda_i^2=1,~~~i=1,...,4.
$$
Comparing $\rho=|\Psi\ra\la\Psi|$ with the general expression (\ref{Pauli}),
we have the following relations after some straightforward calculations,
$$
\sum_{i,j=1}^3(R_{ij}^2+S_{ij}^2+T_{ij}^2)=3,\ \
\sum_{i,j,k=1}^3Q_{ijk}^2+\sum_{i=1}^3(\alpha_i^2+\beta_i^2+\gamma_i^2)=4.
$$
As $\sum\alpha_i^2,\ \sum R_{ij}^2,\ \sum Q_{ijk}^2$ etc. are
all invariants under local unitary transformations \cite{Linden},
the relations above hold not only for state (\ref{Psi4}), but also
for all pure states. We simply denote these relations as
$|\vec{R}|^2+|\vec{S}|^2+|\vec{T}|^2=3$,
$|\vec{Q}|^2+|\vec{\alpha}|^2+|\vec{\beta}|^2+|\vec{\gamma}|^2=4$.
Because $|\vec{\alpha}|$, $|\vec{\beta}|$, $|\vec{\gamma}|\leq1$,
we have $1\leq|\vec{Q}|\leq2$. The minimum is attained when
$|\Psi\ra$ is fully separable and the maximum is obtained for the
maximally entangled state, i.e. the GHZ state.

Set $C_i=\frac{1}{2}(A_i+B_i)$, $D_i=\frac{1}{2}(A_i-B_i)$ and
$\vec{s}_i=\frac{1}{2}(\vec{a}_i+\vec{b}_i)$,
$\vec{t}_i=\frac{1}{2}(\vec{a}_i-\vec{b}_i)$, for $i=1,2,3$. We
have $|\vec{s}_i|^2+|\vec{t}_i|^2=1$ and
$\vec{s}_i\cdot\vec{t}_i=0$. The operators $\mathcal{D}_3^{(i)}$ can be
rewritten as:
\begin{eqnarray*}
\mathcal{D}_3^{(1)}=C_1C_2C_3+C_1C_2D_3+C_1D_2C_3-C_1D_2D_3+D_1,\\
\mathcal{D}_3^{(2)}=C_1C_2C_3+C_1C_2D_3+D_1C_2C_3-D_1C_2D_3+D_2,\\
\mathcal{D}_3^{(3)}=C_1C_2C_3+C_1D_2C_3+D_1C_2C_3-D_1D_2C_3+D_3.
\end{eqnarray*}
Then
\begin{eqnarray*}
\omega&=&\la\Psi|\mathcal{D}_3^{(1)}|\Psi\ra_\rho^2
+\la\Psi|\mathcal{D}_3^{(2)}|\Psi\ra_\rho^2
+\la\Psi|\mathcal{D}_3^{(3)}|\Psi\ra_\rho^2\\
&=&(\vec{s}_1\otimes\vec{s}_2\otimes\vec{s}_3\cdot\vec{Q}
+\vec{s}_1\otimes\vec{s}_2\otimes\vec{t}_3\cdot\vec{Q}
+\vec{s}_1\otimes\vec{t}_2\otimes\vec{s}_3\cdot\vec{Q}
-\vec{s}_1\otimes\vec{t}_2\otimes\vec{t}_3\cdot\vec{Q}
+\vec{t}_1\cdot\vec{\alpha})^2\\
&&+(\vec{s}_1\otimes\vec{s}_2\otimes\vec{s}_3\cdot\vec{Q}
+\vec{s}_1\otimes\vec{s}_2\otimes\vec{t}_3\cdot\vec{Q}
+\vec{t}_1\otimes\vec{s}_2\otimes\vec{s}_3\cdot\vec{Q}
-\vec{t}_1\otimes\vec{s}_2\otimes\vec{t}_3\cdot\vec{Q}
+\vec{t}_2\cdot\vec{\beta})^2\\
&&+(\vec{s}_1\otimes\vec{s}_2\otimes\vec{s}_3\cdot\vec{Q}
+\vec{s}_1\otimes\vec{t}_2\otimes\vec{s}_3\cdot\vec{Q}
+\vec{t}_1\otimes\vec{s}_2\otimes\vec{s}_3\cdot\vec{Q}
-\vec{t}_1\otimes\vec{t}_2\otimes\vec{s}_3\cdot\vec{Q}
+\vec{t}_3\cdot\vec{\gamma})^2,
\end{eqnarray*}
where $\vec{s}\otimes\vec{t}\otimes\vec{p}\cdot\vec{Q}$
stands for $\sum_{ijk} s_i t_j p_k Q_{ijk}$.
From this expression, we see that $\omega$ attains its maximum
at either $|\vec{Q}|=1$ or $|\vec{Q}|=2$. For the
case $|\vec{Q}|=1$, the state $|\Psi\ra$ is fully separable and the inequality is
trivially satisfied. In the second case $|\vec{Q}|=2$ we have:
\begin{eqnarray*}
\omega &=&(\vec{s}_1\otimes\vec{s}_2\otimes\vec{s}_3\cdot\vec{Q}
+\vec{s}_1\otimes\vec{s}_2\otimes\vec{t}_3\cdot\vec{Q}
+\vec{s}_1\otimes\vec{t}_2\otimes\vec{s}_3\cdot\vec{Q}
-\vec{s}_1\otimes\vec{t}_2\otimes\vec{t}_3\cdot\vec{Q})^2\\
&&+(\vec{s}_1\otimes\vec{s}_2\otimes\vec{s}_3\cdot\vec{Q}
+\vec{s}_1\otimes\vec{s}_2\otimes\vec{t}_3\cdot\vec{Q}
+\vec{t}_1\otimes\vec{s}_2\otimes\vec{s}_3\cdot\vec{Q}
-\vec{t}_1\otimes\vec{s}_2\otimes\vec{t}_3\cdot\vec{Q})^2\\
&&+(\vec{s}_1\otimes\vec{s}_2\otimes\vec{s}_3\cdot\vec{Q}
+\vec{s}_1\otimes\vec{t}_2\otimes\vec{s}_3\cdot\vec{Q}
+\vec{t}_1\otimes\vec{s}_2\otimes\vec{s}_3\cdot\vec{Q}
-\vec{t}_1\otimes\vec{t}_2\otimes\vec{s}_3\cdot\vec{Q})^2.
\end{eqnarray*}
Without losing generality, we consider the maximally entangled state
$|\Psi\ra$ of the form
 $|\Psi\ra=\frac{1}{\sqrt{2}}(|000\ra+|111\ra)$. We have then
$Q_{ijk}=0$ except for $Q_{111}=1$,
$Q_{122}=Q_{212}=Q_{221}=-1$. To attain the maximum of $\omega$,
the third components of $\vec{s}_i,\ \vec{t}_i$ should be zero,
and either $|\vec{s}_i|=|\vec{t}_i|$ or one of the $|\vec{s}_i|$ and $|\vec{t}_i|$
is zero and the other one is 1. We deal with these
cases respectively.
\begin{enumerate}
 \item[{(1)}] The cases $|\vec{s}_i|=1,\ |\vec{t}_i|=0$, $i=1,2,3$,
 $$
 \omega=3(\vec{s}_1\otimes\vec{s}_2\otimes\vec{s}_3\cdot\vec{Q})^2
 =3\la\Psi|C_1C_2C_3|\Psi\ra^2\leq3.
 $$
 \item[{(2)}] Cases of the form $|\vec{s}_1|=|\vec{s}_2|=1$ ($|\vec{t}_1|=|\vec{t}_2|=0$ ) and
 $|\vec{s}_3|=|\vec{t}_3|=\frac{1}{\sqrt{2}}$, then
 \begin{eqnarray*}
 \omega &=&2(\vec{s}_1\otimes\vec{s}_2\otimes\vec{s}_3\cdot\vec{Q}
 +\vec{s}_1\otimes\vec{s}_2\otimes\vec{t}_3\cdot\vec{Q})^2
 +(\vec{s}_1\otimes\vec{s}_2\otimes\vec{s}_3\cdot\vec{Q})^2\\[2mm]
 &=&2\la\Psi|C_1C_2(C_3+D_3)|\Psi\ra^2+\la\Psi|C_1C_2C_3|\Psi\ra^2\leq 3.
\end{eqnarray*}
\item[{(3)}] Cases like $|\vec{s}_1|=1$ ($|\vec{t}_1|=0$) and
$|\vec{s}_i|=|\vec{t}_i|=\frac{1}{\sqrt{2}}$ for $i=2,3$. From the
orthogonal relation of $\vec{s}_i$ and $\vec{t}_i$, we can express
them as: $\vec{s}_1=(\mbox{cos}\theta_1,\ \mbox{sin}\theta_1,\ 0)$
and $\vec{s}_i=\frac{1}{\sqrt{2}}(\mbox{cos}\theta_i,\
\mbox{sin}\theta_i,\ 0)$,
$\vec{t}_i=\frac{1}{\sqrt{2}}(-\mbox{sin}\theta_i,\
\mbox{cos}\theta_i,\ 0)$, for $i=2,3$. Direct calculations lead to
\begin{eqnarray}\label{3}
\omega=\frac{3}{2}(\mbox{cos}(\theta_1+\theta_2+\theta_3)
-\mbox{sin}(\theta_1+\theta_2+\theta_3))^2\leq3.
\end{eqnarray}
\item[{(4)}] $|\vec{s}_i|=|\vec{t}_i|=\frac{1}{\sqrt{2}}$.
Set $\vec{s}_i=\frac{1}{\sqrt{2}}(\mbox{cos}\theta_i,\
\mbox{sin}\theta_i,\ 0)$,
$\vec{t}_i=\frac{1}{\sqrt{2}}(-\mbox{sin}\theta_i,\
\mbox{cos}\theta_i,\ 0)$, for $i=1,2,3$. We get the same
result as (\ref{3}).
\end{enumerate}
This ends the proof. \hfill  $\square$

\begin{figure}[tbp]
\begin{center}
\resizebox{8cm}{!}{\includegraphics{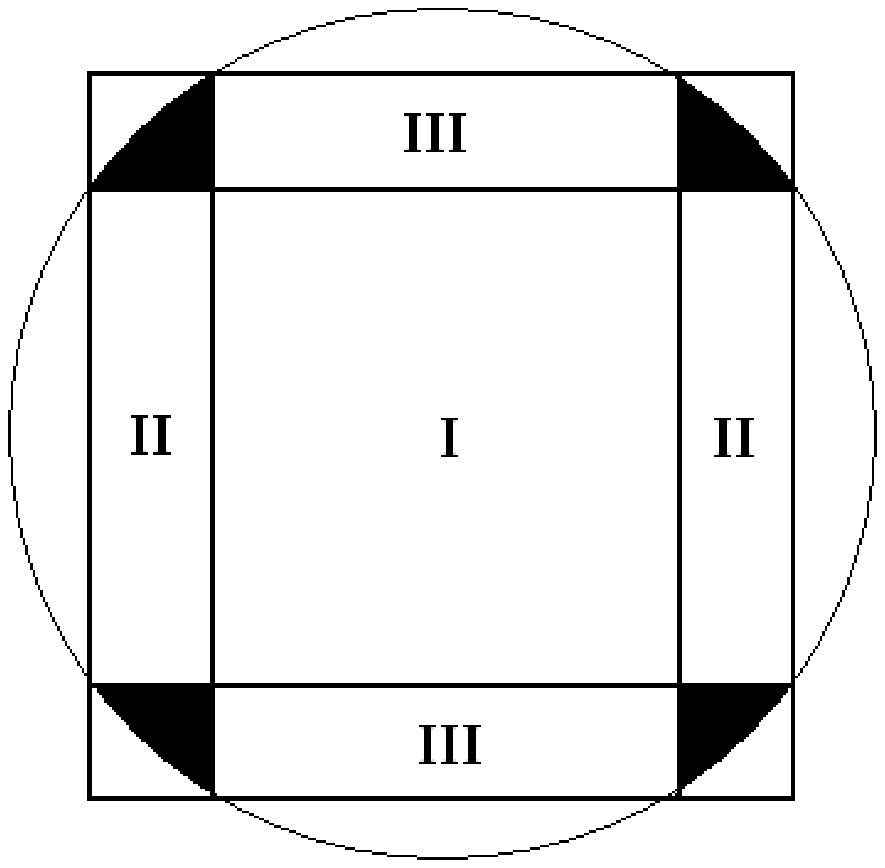}}
\end{center}
\caption{Projection of the sphere related to the space of three-qubit entanglement}
\label{fig1}
\end{figure}

By regarding the average of
$\mathcal{D}_3^{(1)},\mathcal{D}_3^{(2)},\mathcal{D}_3^{(3)}$ as
three axes of the space, we have that all the 3-qubit states
are in the cube of edge length $2\sqrt{2}$ with the center at the
origin. But the inequality in Theorem 2 restricts all the states
into a ball with radius $\sqrt{3}$ centered at the origin, which is
just the external ball of the cube of edge length $2$. Therefore all the
states are located in the common space of the larger cube of edge length $2\sqrt{2}$
and the ball with radius $\sqrt{3}$, where the smaller cube of edge length $2$
is for separable states, and the rest space is for all
kinds of entangled states.

If we project the state space along the direction $\mathcal{D}_3^{(3)}$ to the
plane $(\mathcal{D}_3^{(1)},\mathcal{D}_3^{(2)})$ (cut the sphere
in the $(\mathcal{D}_3^{(1)},\mathcal{D}_3^{(2)})$ plane), we get the
Figure 1. The completely separable states are in the region
labeled I, the left and right
rectangular regions labeled II belong to $S_{1-23}$,
and the top and bottom rectangular regions III belong to
$S_{2-13}$ (the states in $S_{12-3}$ are in front of and
behind the region I which could not be seen due to projection).
The rest entangled states are located in the
(corners) black regions.

We have presented a set of Bell inequalities which distinguish
three possible bi-separable entanglements of tripartite qubit systems.
The three Bell operators we used constitute an external sphere of the
separable cube. By using the Bell operators defined in (\ref{sun}),
the results can be generalized to $N$-qubit system.
We conjecture that, for $N$-qubit system, one would have
$$
\sum_{i=1}^N\la\mathcal{D}_N^{(i)}\ra_\rho^2\leq N.
$$
Nevertheless, the generalized Bell quantities would only detect the entanglement
between one qubit and the rest ones. For instance, for four-qubit systems,
one can only learn from $|\langle\mathcal{D}_4^{(1)}\rangle|>1$ that
the first qubit is entangled with the rest three qubits.
One needs more Bell operators to classify all other possible entanglements.
\bigskip

\noindent We thank Kai Chen, Zeng-Bing Chen, Jian-Wei Pan and Chun-Feng Wu for
valuable discussions and communications. B.Z. Sun gratefully acknowledges the
hospitality of Max-Planck Institute of Mathematics in the Sciences, Leipzig.
The work is partially supported by the Natural Science
Foundation of Shandong Province(Grant No. Y2005A11) and
NSFC 10675086.

\end{document}